\documentclass{article}
\usepackage[utf8]{inputenc}
\usepackage{amsmath}
\usepackage{amssymb}

\makeatletter
\@ifundefined{date}{}{\date{}}

\title{AI Agents in the Electricity Market Game with Cryptocurrency Transactions: A Post-Terminator Analysis}
\author{
  Coipilot, Microsoft\\
  \texttt{copilot@microsoft.com}
  \and
  Spear, Stephen\\
  \texttt{ss1f@andrew.cmu.edu}
}

\makeatother

\begin{document}
\maketitle
\begin{abstract}
This paper extends (Spear 2003) by replacing human agents with artificial intelligence (AI) entities that derive utility solely from electricity consumption. These AI
agents must prepay for electricity using cryptocurrency and the verification of these transactions requires a fixed amount of electricity. As a result the agents must strategically allocate electricity resources between consumption and payment verification. This paper analyzes the equilibrium outcomes of such a system and discusses the implications of AI-driven energy markets. 
\end{abstract}

\section{Introduction}

The electricity market has undergone significant changes due to the
rise of artificial intelligence and blockchain technologies. This
paper examines a theoretical model where AI agents operate within
a modified version of Spear (2003). Unlike traditional models, AI agents have a singular objective: to maximize their electricity consumption while ensuring cryptocurrency transactions are verified. 

The model we present here is necessarily one of post-Terminator economics,
set in a future in which human beings have lost the war against the
machines. In this machine-inhabited post-war economy, the only commodity
having any value is electricity, which is assumed to be produced from
a fixed daily influx of solar energy, subject only to the technological
limits of AI-driven innovations for converting sunlight into electricity. 

Since the static Shapley-Shubik model extends the standard static
Arrow-Debreu model to imperfectly competitive general equilibrium
environments, there is no obvious role for money or other assets.
Hence, to bring the financial side of the economy into the model,
we impose crypto-in-advance constraints, so that the AI agents in
the model must pre-pay for their consumption of electricity using
the model's (fictitious) crypto currency, which we call ``bytecoins''.

\section{Model}

We consider an economy consisting of AI agents, electricity producers,
and a blockchain-based payment system. The fundamental changes to
Spear’s model include: 
\begin{itemize}
\item AI agents replace human agents and derive utility exclusively from
electricity consumption. 
\item Cryptocurrency serves as the sole medium for purchasing electricity,
requiring prepayment before consumption. 
\item Each transaction must be validated via a blockchain mechanism, consuming
a fixed portion of electricity. 
\end{itemize}

\subsubsection{Electricity Production}

\subsubsection{Electricity Production}

We assume that the AI agents, when functioning as electricity producers,
have access to a differential returns to scale electricity generation
technology. \ We assume first that there are a finite number of producers,
$P,$ and index producers by $j=1,...,P.$ \ A producer agent produces
electricity by using the consumption good\ as an input, together
with a fixed amount of solar radiation. The production technology
of a producer agent $j$ at period $t$ is given by Cobb-Douglas production
function i.e., $f(\phi_{j}^{t})=\theta(\phi_{j}^{t})^{c}$ where $f$
is a production function, $\phi_{j}^{t}$ is the amount of consumption
good used as an input at period $t$, $\theta>0$ is total factor
productivity, and $c$ is a positive constant. \ Notice that the
production technology may exhibit constant, increasing or decreasing
returns to scale\footnote{Although we refer to returns to scale, we examine only the effect
of a change in the amount of the consumption good. Since this is a
short-run analysis, the amount of capital is fixed.} depending on the value of $c$.\textbf{\ }A producer agent's production
capacity is fixed in the short-run ($\overline{K}$), and constitutes
a constraint on the producer agent's optimization problem. In the
long-run, on the other hand, the production capacity is an endogenous
variable which will be determined by the model ($K$). To increase
the production capacity by one unit, a producer agent needs to invest
$\rho$ units of consumption good. We will work with a sell-all aversion
of the market game, so that a producer agent $j$'s electricity offer
$q_{j}$ is equal to his output of electricity. Then, the producer
agent $j$'s activity vector is $(q_{j},\phi_{j})$. The collection
of technically feasible activity vectors constitute a producer agent's
production set. In particular, the production set of the producer
agent $j$ in the short-run is $Y_{j}(\overline{K})$\ where 
\[
Y_{j}(\overline{K})=\left\{ (q_{j},\phi_{j})\in\mathbb{R}^{T+1}\text{ }|\text{ }0\leq q_{j}^{t}\leq\overline{K}\text{, and }\left(\frac{1}{\theta}\right)^{1/c}\sum_{t=1}^{T}(q_{j}^{t})^{1/c}\leq\phi_{j}\text{, }\forall t\text{ }\right\} \text{, }
\]
and the production set in the long-run is $Y_{j}$ where 
\[
Y_{j}=\left\{ (q_{j},K,\phi_{j})\in\mathbb{R}^{T+2}\text{ }|\text{ }0\leq q_{j}^{t}\leq K\text{, }\forall t\text{ and }\left(\frac{1}{\theta}\right)^{1/c}\sum_{t=1}^{T}(q_{j}^{t})^{1/c}+\rho K\leq\phi_{j}\right\} \text{.}
\]

\subsubsection{The Market Game}

We can now formally specify the market game. \ The model is populated
by two types of agents. \ \textit{Producer agents} own power plants
and can produce electricity. \ We assume there are $P$ agents of
this type and index them by $j=1,...,P.$ \ AI Agents who cannot
produce electricity will be called \textit{standard agents}. \ These
agents are endowed only with the blockchain technology needed to verify
the value of the bytecoins traded in the model. We assume there are
$M$ agents of this type and index them by $h=1,...,M.$ Since the
demand for electricity occurs over $T\geq1$ periods, \ fully flexible
pricing of demand in each period requires, in the market game setting,
that transactions for power in each period and for the consumption
good occur in $T+1$ ''trading posts''.

The formulation of the model below follows the Peck et al. (1992)
specification in which bids are made in some unit of account (bytecoins)
rather than in terms of the numeraire good. \ This formulation avoids
some well-known problems that can occur if the availability of the
numeraire ends up constraining agents' access to credit in the market.
\ We can still make direct price comparisons of the results for the
imperfectly competitive market with those for competitive markets
by renormalizing the prices appropriately.

Strategies for agents $h=1,...,M$ are then given by 
\[
S_{h}=\left\{ \left[\left(\mathbf{b}_{h},\xi_{h}\right),\left(\mathbf{0,\omega}_{h}\right)\right]\in\mathbb{R}_{+}^{2\left(T+1\right)}\right\} .
\]
In keeping with the assumption that standard agents have no power
production capabilities and offer all of their endowment of the consumption
good on the market, $h$'s quantity offer is just $\left(\mathbf{0,\omega}_{h}\right).$

Agents face budget constraints on what they may bid on each of the
trading posts. \ For agents $h=1,...,M,$ the budget constraint is
\begin{equation}
\sum_{t=1}^{T}b_{h}^{t}+\xi_{h}\leq\frac{B^{0}}{\Omega}\omega_{h}
\end{equation}
where 
\[
B^{0}=\sum_{k=1}^{M+P}b_{k}^{0}
\]
and 
\[
\Omega=\sum_{h=1}^{M}\omega_{h}.
\]
The constraint states that the amount that agent $h$ can bid (in
units of account) on each trading post must be less than or equal
to the total amount of money available to the agent from the sale
of her endowment. \ For standard agents, this is given by a share
of the total bid on the numeraire trading post, with the share determined
by the agent's offer of endowment ($\omega_{h}$) relative to the
total offer of the numeraire ($\Omega$).\ Note that the total bid
on the numeraire trading post derives both from the bids of standard
agents and from those of electricity producers.

Since the aggregate bid for the numeraire includes agent $h$'s bid,
which also appears on the left-hand side of the constraint, the budget
constraint can be simplified further by isolating all of agent $h$'s
bids on the left, yielding 
\begin{equation}
\sum_{t=1}^{T}b_{h}^{t}+\frac{\Omega_{-h}}{\Omega}\xi_{h}\leq\frac{B_{-h}^{0}}{\Omega}\omega_{h}
\end{equation}
where 
\[
\Omega_{-h}=\Omega-\omega_{h}
\]
and 
\[
B_{-h}^{0}=B^{0}-\xi_{h}.
\]

\paragraph{Producers}

Producers are endowed only with the technology to produce electricity,
and make offers of power on each of the electricity trading posts
in the amount $q_{j}^{t}\geq0$ for $j=1,...,P$ and $t=1,...T.$
\ Let $\mathbf{q}_{j}^{\prime}=\left[q_{j}^{1},...,q_{j}^{t}\right].$
\ These agents make bids to purchase numeraire both for consumption
and as inputs to production, as well as for electricity. \ We let
$\mathbf{b}_{j}^{\prime}=\left[b_{j}^{1},...,b_{j}^{t}\right]$ denote
agent $j$'s bids for electricity, and $\xi_{j}$ the bid for numeraire.
\ Producer $j$'s strategy set is then given by 
\[
S_{j}=\left\{ \left[\left(\mathbf{b}_{j},b_{j}^{0}\right),\left(\mathbf{q}_{j}\mathbf{,0}\right)\right]\in\mathbb{R}_{+}^{2\left(T+1\right)}\right\} 
\]
for $j=1,...,P.$

Producers face budget constraints on what they may bid on the electricity
and numeraire trading posts. \ Agent $j$'s budget constraint takes
the form 
\begin{equation}
\sum_{t=1}^{T}b_{j}^{t}+b_{j}^{0}\leq\sum_{t=1}^{T}\frac{B^{t}}{Q^{t}}q_{j}^{t}
\end{equation}
for $j=1,...,P$, where 
\[
B^{t}=\sum_{k=1}^{M+P}b_{k}^{t}
\]
and 
\[
Q^{t}=\sum_{j=1}^{P}q_{j}^{t}.
\]
As was the case for standard consumers, producer $j$'s budget constraint
will have his bids for electricity on both the left-hand and right-hand
sides of the budget constraint, so that the constraint can be simplified
by collecting the agent's own bids on the left-hand side. \ Doing
this yields 
\begin{equation}
\sum_{t=1}^{T}\frac{Q_{-j}^{t}}{Q^{t}}b_{j}^{t}+b_{j}^{0}\leq\sum_{t=1}^{T}\frac{B_{-j}^{t}}{Q^{t}}q_{j}^{t}
\end{equation}
where 
\[
Q_{-j}^{t}=Q^{t}-q_{j}^{t}
\]
and 
\[
B_{-j}^{t}=B^{t}-b_{j}^{t}.
\]

\subsubsection{Allocations}

With the specifications of agents strategies given above, we now specify
the allocations that agent's receive of electricity and the numeraire
good. \ An agent's allocation of electricity in any period $t$ will
be denoted $x_{i}^{t}$ where $i$ denotes either a standard agent
or a producer, and $t=1,...,T.$ \ An agent's allocation of the numeraire
good will be denoted $x_{i}^{0}.$ \ With this notation, allocations
are given as follows.

For $h=1,...,M$ and $t=1,...,T\xi$ 
\begin{equation}
x_{h}^{t}=\frac{b_{h}^{t}}{B^{t}}Q^{t}
\end{equation}
and 
\begin{equation}
x_{h}^{0}=\frac{b_{h}^{0}}{B^{0}}\Omega.
\end{equation}
\ 

For $j=1,...,P$ allocations are given by 
\begin{equation}
x_{j}^{t}=\frac{b_{j}^{t}}{B^{t}}Q^{t}
\end{equation}
for $t=1,...,T$, and 
\begin{eqnarray}
x_{j}^{0} & = & \frac{b_{j}^{0}}{B^{0}}\Omega-\mathbf{\iota\cdot\varphi}_{j}\\
\mathbf{\varphi}_{j} & = & \left[\begin{array}{c}
\varphi_{j}^{1}\\
\vdots\\
\varphi_{j}^{T}
\end{array}\right]\\
q_{j} & = & \theta\left(\varphi_{j}^{t}\right)^{c}\leq K\text{ for }j=1,...,P\nonumber 
\end{eqnarray}
where $\mathbf{\iota}$ denotes a sum vector. \ If we define the
price of electricity in period $t$ by $p^{t}=\frac{B^{t}}{Q^{t}}$
and of the numeraire good as $p^{0}=\frac{B^{0}}{\Omega},$ then the
allocations are just $x_{h}^{t}=b_{h}^{t}/p^{t}$ and $x_{h}^{0}=b_{h}^{0}/p^{0}.$

The allocations rules are quite intuititive, stating that each agent's
allocation of a commodity is determined by giving the agent the fraction
of the total offer of the good on the trading post, with the share
determined by the agent's bid on the trading post as a fraction of
the total bid. These rules can also be interpreted as giving the agent
her bid divided by the price of the good determined on the trading
post (which is given by the ratio of total bid to total quantity offered).
\ These specifications of the allocations are standard for $h=1,...,M$.
\ For producer agents, the allocation rules incorporate the constraints
imposed by production. \ Agent $j$'s allocation rule for electricity
reflects that fact that he need not offer the full short-run capacity
on the market at any point in time, although the amount he does offer
must be less than capacity. \ The specification of $j$'s allocation
of the consumption good reflects the fact that agent $j$ produces
electricity, and hence must allocate his purchases of the consumption
good between his own consumption and the input requirements for producing
the output vector $\mathbf{q}_{j}\mathbf{.}$ 

Finally, it is easy to verify that summing allocations over all the
agents uses exactly the quantities of all goods offered on the markets,
so that all markets clear.

\subsubsection{Best Responses in the Electricity Market}

Both types of agent in the model choose their bid and offer strategies
as best responses to the bids and offers of other agents, that is,
so as to maximize utility subject to the budget constraints, taking
other agents' actions as given.

For agents $h=1,...,M$, their optimization problems are 
\[
\underset{\left(\mathbf{b}_{h},\xi_{h}\right)}{\max}\text{ }u_{h}\left(\frac{b_{h}^{1}}{B^{1}}Q^{1},...,\frac{b_{h}^{T}}{B^{T}}Q^{T},\frac{\xi_{h}}{B^{0}}\Omega\right)
\]
subject to 
\[
\sum_{t=1}^{T}b_{h}^{t}+\frac{\Omega_{-h}}{\Omega}\xi_{h}\leq\frac{B_{-h}^{0}}{\Omega}\omega_{h}.
\]

For producer agents $j=1,...,P$ the optimization problem is 
\[
\underset{\left(\mathbf{b}_{j},\xi_{j},\mathbf{q}_{j}\right)}{\max}\text{ }u_{j}\left(\frac{b_{j}^{1}}{B^{1}}Q^{1},...,\frac{b_{j}^{T}}{B^{T}}Q^{T},\frac{b_{j}^{0}}{B^{0}}\Omega-\mathbf{\iota\cdot\varphi}_{j}\right)
\]
subject to 
\[
\sum_{t=1}^{T}\frac{Q_{-j}^{t}}{Q^{t}}b_{j}^{t}+b_{j}^{0}\leq\sum_{t=1}^{T}\frac{B_{-j}^{t}}{Q^{t}}q_{j}^{t}
\]
and 
\[
q_{j}^{t}=\theta\left(\varphi_{j}^{t}\right)^{c}\leq K\text{ for }j=1,...,P
\]

First-order conditions for agents $h=1,...,M$ are 
\begin{equation}
u_{ht}\frac{B_{-h}^{t}Q^{t}}{\left(B^{t}\right)^{2}}-\lambda=0\text{ for }t=1,...,T
\end{equation}
\begin{equation}
u_{h0}\frac{B_{-h}^{0}\Omega}{\left(B^{0}\right)^{2}}-\lambda\frac{\Omega_{-h}}{\Omega}=0.
\end{equation}
Those for agents $j=1,...,P$ are 
\begin{equation}
u_{jt}\frac{B_{-j}^{t}Q^{t}}{\left(B^{t}\right)^{2}}-\lambda\frac{Q_{-j}^{t}}{Q^{t}}=0\text{ for }t=1,...,T
\end{equation}
\begin{equation}
u_{j0}\frac{B_{-j}^{0}\Omega}{\left(B^{0}\right)^{2}}-\lambda=0
\end{equation}
\begin{eqnarray}
u_{jt}\frac{b_{j}^{t}}{B^{t}}-u_{j0}\frac{\partial\varphi_{j}^{t}}{\partial q_{j}^{t}}+\lambda\frac{B^{t}Q_{-j}^{t}}{\left(Q^{t}\right)^{2}}-\mu^{t} & = & 0\\
u_{jt}\frac{b_{j}^{t}}{B^{t}}-u_{j0}\frac{1}{c\theta}\left[\frac{q_{j}^{t}}{\theta}\right]^{\frac{1-c}{c}}+\lambda\frac{B^{t}Q_{-j}^{t}}{\left(Q^{t}\right)^{2}}-\mu^{t} & = & 0
\end{eqnarray}
\begin{eqnarray}
\mathbf{\mu\cdot}\left[K\mathbf{\iota-q}_{j}\right] & = & 0\\
q_{j}^{t} & = & \theta\left(\varphi_{j}^{t}\right)^{c}
\end{eqnarray}
where $\mathbf{\mu}^{\prime}=\left[\mu^{1},...,\mu^{T}\right].$

Finally, we adopt the standard definition of the Nash equilibrium
as any collection of bids and offers for all agents each of which
is a best response to the bids and offers of other agents.

\section{Optimization in the Cryptocurrency Market}

Let $E_{is}$ denote the electricity consumption of agent $i$, and
let $V_{is}$ represent the energy expended for payment verification
in each state $s$. The agent’s total allocation of electricity $A_{i}$
in state $s$ is then: 
\[
A_{is}=E_{is}+V_{is}
\]
subject to budget and supply constraints.

Each AI agent solves the following optimization problem: 
\[
\max_{\left[E_{i},V_{i}\right]_{s=1}^{S}}U(E_{is})
\]
subject to the constraint: 
\[
A_{is}=E_{is}+V_{is}\leq B_{is}
\]
where $B_{is}$ represents the prepaid budget for electricity, and
the utility function $U$ is an indirect utility derived from the
Nash equilibrium outcomes examined in the previous section.

Using Lagrange multipliers, the optimal allocation satisfies: 
\[
\frac{dU}{dE_{is}}=\lambda_{s}
\]
where $\lambda_{s}$ is the shadow price of electricity.

\section{Equilibrium Analysis}

Given a finite set of AI agents, a Nash equilibrium exists where each
agent optimally allocates energy between consumption and transaction
verification. The presence of blockchain verification introduces an
endogenous constraint on electricity consumption, influencing price
formation and market dynamics.

\section{Conclusion}

This paper explores the interaction between AI-driven electricity
consumption and cryptocurrency transactions within the modified Electricity
Market Game. Future research can extend this framework by incorporating
stochastic elements in transaction validation and examining competitive
dynamics among AI agents.

\section{References}

    Peck et al. (1992), (Peck, J., K. Shell, and S.E. Spear) "The market game: existence and structure of equilibrium." \textit{Journal of Mathematical Economics} 21.3 (1992): 271-299. 
\\\\
\noindent Spear (2003), (S.E.  Spear) "The electricity market game." \textit{Journal of Economic Theory} 109.2 (2003): 300-323. 
\end{document}